\documentclass[preprint, aps, pra, showpacs]{revtex4-1}
\usepackage{eurosym}
\usepackage{xcolor}
\usepackage{amssymb}
\usepackage{graphicx}
\usepackage{dcolumn}
\usepackage{bm}
\usepackage{amsmath}

\begin{document}

\title{Moir\'e patterns in doubly differential electron momentum
distributions in atomic ionization by midinfrared lasers}
\author{Mart\'{\i}n Dran}
\affiliation{Institute for Astronomy and Space Physics IAFE (UBA-Conicet), Buenos Aires,
Argentina}
\author{Diego G. Arb\'o}
\affiliation{Institute for Astronomy and Space Physics IAFE (UBA-Conicet), Buenos Aires,
Argentina}
\date{\today }

\begin{abstract}
We analyze the doubly differential electron momentum distribution in
above-threshold ionization of atomic hydrogen by a linearly-polarized
mid-infrared laser pulse. We reproduce side rings in the momentum
distribution with forward-backward symmetry previously observed by Lemell 
\textit{et al.} in Phys. Rev. A \textbf{87}, 013421(2013), whose origin, as
far as we know, has not been explained so far. By developing a Fourier
theory of moir\'{e} patterns, we demonstrate that such structures stems from
the interplay between intra- and intercycle interference patterns which work
as two separate grids in the two-dimensional momentum domain. We use
a three dimensional (3D) description based on the saddle-point approximation
(SPA) to unravel the nature of these structures. When the
periods of the two grids (intra- and intercycle) are similar, principal moir%
\'{e} patterns arise as concentric rings symmetrically in the forward and backward
directions at high electron kinetic energy. Higher order moir\'{e}
patterns are observed and characterized when the period of one grid is
multiple of the other. We find a scale law for the position (in momentum
space) of the center of the moir\'{e} rings in the tunneling regime. 
We verify the SPA predictions by comparison with time-dependent distorted 
wave strong-field approximation (SFA) calculations and the solutions of the full 3D
time-dependent Schr\"{o}dinger equation (TDSE).
\end{abstract}

\maketitle
\preprint{APS/123-QED}


\section{\label{sec:level1}Introduction}


In a typical photoionization process in the tunneling regime, electrons are
emitted by tunneling through the potential barrier formed by the combination
of the atomic potential and the external strong field. Tunneling occurs
within each optical cycle predominantly around the maxima of the absolute
value of the electric field. According to the well-known three-step model,
photoelectrons can be classified into \textit{direct} and \textit{rescattered%
} electrons \cite{Kulander95,Paulus94,Lewenstein95}. After ionization,
direct electrons can escape without being strongly affected by the residual
core potential. The classical cutoff energy for this process is twice the
ponderomotive energy. After being accelerated back by the laser field, a
small portion of electrons are rescattered by the parent ion and can achieve
a kinetic energy of up to ten times de ponderomotive energy. Trajectories
that correspond to direct ionization are crucial in the formation of
interference patterns in photoelectron spectra. Quantum interference within
an optical cycle was firstly reported (as far as we know) in Ref. \cite%
{Gribakin97} and theoretically analyzed and experimentally observed by
Paulus \textit{et al} in \cite{Paulus98} both for negative ions. A thorough
saddle-point analysis with the strong field approximation can be found in
Becker's review \cite{Becker02}. Non-equidistant peaks in the photoelectron
spectrum were firstly calculated for neutral atoms by Chirila \textit{et al} 
\cite{Chirila05}. A temporal double-slit interference pattern has been
studied in near-single cycle pulses both experimentally \cite%
{Lindner05,Gopal09} and theoretically \cite{Becker02,Arbo06b}. Near
threshold oscillations in angular distribution were explained as
interferences of electron trajectories \cite{Arbo06a} and measured
by \cite{Marchenko10}. Diffraction fringes have been experimentally observed
in photoionization of He \cite{Gopal09,Xie12} and Ne atoms \cite{Xie12} and
photodetachment in H$^{-}$ and \cite{Reichle01} F$^{-}$ ions by femtosecond
pulses for fixed frequency \cite{Bergues07} and theoretically analyzed \cite%
{Bivona08,Arbo12,Arbo10a,Arbo10b}. Diffraction patterns were also found in
spectra of laser-assisted XUV ionization, whose gross structure of sidebands
were explained as the interference between electrons emitted within one
period \cite{Kazansky10a,Bivona10,Gramajo16,Gramajo17,Gramajo18}. The
interference pattern in multi-cycle photoelectron spectra can be identified
as a diffraction pattern at a time grating composed of intra- and intercycle
interferences \cite{Bivona08,Arbo10a,Arbo10b,Arbo12}. While the latter gives
rise to the well-known ATI peaks \cite{Agostini79,Protopapas97,Joachain00},
the former leads to a modulation of the ATI spectrum in the near infrered regime
offering information on the subcycle ionization dynamics.

In previous papers we analyzed how the interplay between the intercycle
interference [factor $B(k)$ in Eq. (\ref{eq:24})] and the intracycle
interference [factor $F(\vec{k})$ in Eq. (\ref{eq:24})] controls the doubly
differential distribution of direct ATI electrons for lasers in the NIR \cite%
{Arbo10a,Arbo10b,Arbo12}.
In a theoretical study about the quantum-classical correspondence in atomic ionization
by midinfrared pulses, Lemell \textit{et al.} calculated the doubly
differential momentum distribution after the interaction of a strong
midinfrared laser pulse with a hydrogen atom, which shows multiple peaks and
interference structures (see Fig. 1 of \cite{Lemell13}). At both sides of
the well-known intercycle ATI rings, two distinct ring-like structures
appear (symmetrically) in the forward and backward directions. As far as we
know, the origin of these structures has not been identified so far.
In this paper, we extend the analysis of the SPA to the midinfrared regime. 

Large scale interference~patterns can be produced when a small scale grid is overlaid
on another similar grid \cite{Lebanon01,Miao16}. These patterns are named moir\'e  
\cite{Lebanon01,Miao16} and appear in art, physics, mathematics, etc.. 
They show up in everyday life such as a striped shirt seen on television,
in the folds of a moving curtain, when looking through parallel wire-mesh fences, etc.
More than a rareness, moir\'{e} is widely used in projection interferometry
complementing conventional holographic interferometry, especially for
testing optics used at long wavelength. The use of moir\'{e} for reduced
sensitivity testing was introduced by Lord Rayleigh in 1874 to determine the
quality of two identical gratings even though each individual grating could
not be resolved under a microscope \cite{Rayleigh}. Moir\'{e} patterns have
been extremely useful to help the understanding of basic interferometry and
interferometric test results \cite{Oster63,Oster64,Nishijima64}.

In the present communication, we theoretically investigate on the origin of
side ring structures that appear in the doubly differential momentum
distribution for atomic ionization by laser pulses in the midinfrared
spectral region \cite{Lemell13}. We demonstrate that such structures stems
from the interplay between intra- and intercycle interference patterns which
work as two separate grids in the two-dimensional momentum domain. When the
periods of the two grids (intra- and intercycle) are similar, principal moir%
\'{e} patterns arise as concentric rings at high electron kinetic energy in
the forward and backward directions symmetrically. Besides, we show that a
whole family of secondary moir\'{e} patterns with less visibility of the
principal one is also present. We characterize these structures within the
Fourier theory of the moir\'{e} patterns finding simple scale laws for the
position of their center in the momentum distribution. In order to do that,
we previously discard the formation of spurious (non-physical) moir\'{e}
patterns due to the presence of the numerical grid of the momentum map. We
use a three dimensional (3D) description based on the saddle-point
approximation (SPA) \cite{Arbo10a,Arbo10b,Arbo12} to unravel the nature of
these structures. Our SPA predictions are corroborated by comparison with
time-dependent distorted wave strong-field approximation (SFA)\cite%
{Lewenstein95,Chirila05,Macri03,Rodriguez04,Ivanov95} calculations and the
solutions of the full time-dependent Schr\"{o}dinger equation (TDSE).

The paper is organized as follows. In the first part of Sec.\ II A, 
we develop the Fourier theory of moir\'{e} patterns. We continue
by scheming the semiclassical model for atomic ionization by laser pulses showing that
the separation of intracycle and intercycle interferences can be interpreted
in terms of diffraction at a time grating when studying the doubly
differential distributions within the SPA. In the last part of the section
we show how moir\'e patterns are formed from inter- and intracycle interferences
in view of this Fourier theory. In Sec.\ III, we analyze the ring-like structures
in the doubly differential momentum distribution within the SPA
and compare with the SFA and TDSE \textit{ab initio} calculations. We also
characterize the moir\'{e} structure by analyzing the dependence of the
position of the center as a function of laser parameters finding a scale
law. Atomic units are used throughout the paper, except when otherwise
stated.


\section{\label{sec:level2}Theory}

\subsection{Fourier theory of moir\'{e} patterns}

We define a 1D grating (vertical stripes) as a periodic function $%
G(x^{\prime })$, of period $p$ is the period of the grating.
Due to its periodicity, the function $G(x^{\prime })$ can be thought
as a sum of different harmonic terms of discrete frequency,
\begin{equation}
G(x^{\prime }) = \sum_{n=-\infty }^{\infty }a_{n}\exp [i2\pi nf_{0}x^{\prime }],
\label{1D-grating}
\end{equation}%
where $a_{n}$ is the Fourier coefficient and $f_{0}=p^{-1}.$

Gratings with a general geometrical layout can be considered as extended
coordinate-transformed structures which can be obtained by applying
geometric transformations to a standard 1D-grating. By replacing $x^{\prime
} $ with a certain function $T(x,y),$ the 1D grating of Eq. (\ref{1D-grating})]
can be transformed into another curvilinear grating $G_{T}(x,y)=G[T(x,y)]$.
Therefore, in the same way, the latter can be expressed as%
\begin{equation}
G_{T}(x,y)=\sum_{n=-\infty }^{\infty }a_{n}\exp \left[ i2\pi nf_{0}T(x,y)%
\right] .  \label{2D-grating}
\end{equation}

Moir\'{e} fringes appear in the overlay of repetitive structures and vary in
terms of the geometrical layout of two (or more) superposed structures. The
two gratings with the extended layout can be obtained by applying the
transformations $T_{1}(x,y)$ and $T_{2}(x,y)$ to two 1D gratings of
frequencies $f_{1}$ and $f_{2}$, respectively. The generalized gratings can
be expressed as in Eq. (\ref{2D-grating}), 
\begin{subequations}
\begin{eqnarray}
G_{1}(x,y) &=&\sum_{n=-\infty }^{\infty }a_{n}\exp [i2\pi nf_{1}T_{1}(x,y)],
\label{G1} \\
G_{2}(x,y) &=&\sum_{m=-\infty }^{\infty }b_{m}\exp [i2\pi mf_{2}T_{2}(x,y)].
\label{G2}
\end{eqnarray}

The two superimposed gratings can be written as the multiplication of the
two general gratings $G_{1}$ and $G_{2},$ in respective equations (\ref{G1})
and (\ref{G2}), 
\end{subequations}
\begin{eqnarray}
G(x,y) &=&G_{1}(x,y)G_{2}(x,y)  \label{G} \\
&=&\sum_{n=-\infty }^{\infty }\sum_{m=-\infty }^{\infty }a_{n}b_{m}\exp
\left\{ i2\pi \left[ nf_{1}T_{1}(x,y)+mf_{2}T_{2}(x,y)\right] \right\} . 
\notag
\end{eqnarray}%
From Eq. (\ref{G}), we can extract the partial sum $\sum_{n=-\infty
}^{\infty }\sum_{m=-\infty }^{\infty }a_{n}b_{m}(\cdots)\rightarrow \sum_{j=-\infty
}^{\infty }a_{jk_{1}}b_{jk_{2}}(\cdots)$, with $k_{1}$ and $k_{2}$ integer numbers
different from zero. Then, we express this partial sum in the same way as in
Eq. (\ref{2D-grating}), namely,%
\begin{equation}
\tilde{G}(x,y)=\sum_{j=-\infty }^{\infty }a_{jk_{1}}b_{jk_{2}}\exp \left\{
i2\pi jf\left[ k_{1}\left( f_{1}/f\right) T_{1}(x,y)+k_{2}\left(
f_{2}/f\right) T_{2}(x,y)\right] \right\} ,  \label{Gtilde2D}
\end{equation}%
where $f$ is a standardized frequency. In this way, the twofold sum of Eq. (%
\ref{G}) can be decomposed into many partial sums. Eq. (\ref{Gtilde2D}) can
be regarded as the transformation of a 1D grating with a compound
transformation function%
\begin{equation}
T(x,y)=k_{1}\left( \frac{f_{1}}{f}\right) T_{1}(x,y)+k_{2}\left( \frac{f_{2}%
}{f}\right) T_{2}(x,y),  \label{Txy}
\end{equation}%
applied to the 1D grating%
\begin{equation}
\tilde{G}(x^{\prime })=\sum_{j=-\infty }^{\infty }a_{jk_{1}}b_{jk_{2}}\exp
\left( i2\pi jfx^{\prime }\right) .  \label{Gtilde1D}
\end{equation}%
For every pair $(k_{1},k_{2}),$ the partial sum in Eq. (\ref{Gtilde1D})
converges to a periodic-distributed pattern similar to the layout of
standard 1D gratings. By transforming the partial sum of Eq. (\ref{Gtilde1D}%
) with the transformation function of Eq. (\ref{Txy}), we get the $%
(k_{1},k_{2})$-order moir\'{e} pattern of Eq. (\ref{Gtilde2D}). Summing up,
we can say that two geometrically transformed 1D gratings exhibit equivalent
patterns to the one obtained by application of a compound transformation to
a certain 1D-distributed moir\'{e} pattern.

In general, moir\'{e} fringes generated by two superposed gratings are
transformed from two standard 1D gratings with different frequencies by
different transformations. However, in the following section, we restrict to
the special case of moir\'{e} fringes generated from 1D gratings with the
same frequency, i.e., $f_{1}=f_{2}=f$, and different transformations, i.e., $%
T_{1}(x,y)\neq T_{2}(x,y).$ Therefore, the moir\'{e} pattern of Eq. (\ref%
{Gtilde2D}) can be written as 
\begin{equation}
\tilde{G}(x,y)=\sum_{j=-\infty }^{\infty }a_{jk_{1}}b_{jk_{2}}\exp \left\{
i2\pi jf\left[ k_{1}T_{1}(x,y)+k_{2}T_{2}(x,y)\right] \right\} .
\label{moirek1k2}
\end{equation}%
The lowest frequency pattern corresponds to the pair $(k_{1},k_{2})=(1,-1)$,
which is usually the most visible one. We name the pair $(1,-1)$ as the
principal moir\'{e} pattern with the transformation $%
T(x,y)=T_{1}(x,y)-T_{2}(x,y)$. Higher order or secondary moir\'{e} patterns
are also present with less visibility. Later, we will see how the
side-ring structure can be thought of as the principal moir\'{e} pattern
arising from the superposition of intra- and intercycle interferences, each
considered as a separate grid $G_{1}$ and $G_{2}$. But before, in the
next subsection, we pose the semiclassical theory of inter- and intracycle
interference in the electron yield after atomic ionization by a short
laser pulse.

\subsection{Semiclassical model}


In this subsection we repeat the theory of the semiclassical model for
atomic ionization in the single active electron approximation interacting
with a linearly polarized laser field $\vec{F}(t)$ firstly posed in \cite%
{Arbo12,Arbo10a,Arbo10b}. The reader familiar with the semiclassical model
can skip this subsection and go directly to the analysis of the formation of
the moir\'{e} patterns in the next subsection.

The Hamiltonian of the system in the length gauge is 
\begin{equation}
H=\frac{\vec{p}{\,}^{2}}{2}+V(r)+\vec{r}\cdot \vec{F}\,(t),  \label{hami}
\end{equation}%
where $V(r)$ is the atomic central potential and $\vec{p}$ and $\vec{r}$ are
the momentum and position of the electron, respectively. The term $\vec{r}%
\cdot \vec{F}\,(t)$ couples the initial state $|\phi _{i}\rangle $ to the
continuum final state $|\phi _{f}\rangle $ with momentum $\vec{k}$ and
energy $E=k^{2}/2$. The TDSE for the Hamiltonian of Eq. (\ref{hami}) governs
the evolution of the electronic state $\left\vert \psi (t)\right\rangle $.
We calculate the photoelectron momentum distributions as 
\begin{equation}
\frac{dP}{d\vec{k}}\mathbf{=}\left\vert T_{if}\right\vert ^{2},  \label{P}
\end{equation}%
where $T_{if}$ is the T-matrix element corresponding to the transition $\phi
_{i}\rightarrow \phi _{f}$.

The transition amplitude within the time-dependent distorted wave theory in
the strong field approximation (SFA) in the \textit{post} form is expressed
as \cite{Dewangan94}

\begin{equation}
T_{if}=-i\int\limits_{-\infty }^{+\infty }dt\ \langle \chi
_{f}^{-}(t)|z\,F\,(t)\left\vert \phi _{i}(t)\right\rangle ,  \label{Tif}
\end{equation}%
where $\chi _{f}^{-}(t)$ is the final distorted-wave function and the
initial state $\phi _{i}(t)$ is an eigenstate of the atomic Hamiltonian
without perturbation with eigenenergy equal to minus the ionization
potential $I_{p}$. If we choose the Hamiltonian of a free electron in the
time-dependent electric field as the exit-channel distorted Hamiltonian,
i.e., $i\frac{\partial }{\partial t}\left\vert \chi _{f}^{-}(t)\right\rangle
=\left( \frac{p^{2}}{2}+z\,F\,(t)\right) |\chi _{f}^{-}(t)\rangle \,,$ the
solutions are the Volkov states \cite{Volkov35} 
\begin{equation}
\chi _{\vec{k}}^{(V)-}(\vec{r},t)=\frac{\exp \mathbf{[}i(\vec{k}+\vec{A}%
)\cdot \vec{r}\mathbf{]}}{\left( 2\pi \right) ^{3/2}}\exp \left[ iS(t)\right]
\ ,  \label{V-dwf}
\end{equation}%
where $S$ denotes the Volkov action 
\begin{equation}
S(t)=-\int_{t}^{\infty }dt^{\prime }\left[ \frac{(\vec{k}+\vec{A}(t^{\prime
}))^{2}}{2}+I_{p}\right] .  \label{action}
\end{equation}%
In equations (\ref{V-dwf}) and (\ref{action}), $\vec{A}(t)=-\int_{-\infty
}^{t}dt^{\prime }\vec{F}(t^{\prime })$ is the vector potential of the laser
field divided by the speed of light. Eq.\ (\ref{Tif}) together with Eq.\ (%
\ref{V-dwf}) leads to the SFA transition matrix. Accordingly, the influence
of the atomic core potential on the continuum state of the receding electron
is neglected and, therefore, the momentum distribution is a constant of
motion after conclusion of the laser pulse \cite{Lewenstein95,Arbo08a}.

To solve the time integral in Eq. (\ref{Tif}), we closely follow the
\textquotedblleft saddle-point approximation\textquotedblright\ (SPA) \cite%
{Chirila05,Corkum94,Ivanov95,Lewenstein95}, which considers the transition
amplitude as a coherent superposition of electron trajectories 
\begin{equation}
T_{if}(\vec{k})=-\sum_{i=1}^{M}G(t_{r}^{(i)},\vec{k})\ e^{iS(t_{r}^{(i)})}.
\label{bsol}
\end{equation}%
Here, $M$ is the number of trajectories born at ionization times $%
t_{r}^{(i)} $ reaching a given final momentum $\vec{k}$, and $G(t_{r}^{(i)},%
\vec{k})$ is the ionization amplitude, 
\begin{equation}
G(t_{r}^{(i)},\vec{k})=\left[ \frac{2\pi iF(t_{r}^{(i)})}{|\vec{k}+\vec{A}%
(t_{r}^{(i)})|}\right] ^{1/2}d^{\ast }\left( \vec{k}+\vec{A}\left(
t_{r}^{(i)}\right) \right) ,  \label{eq:11}
\end{equation}%
where $d^{\ast }(\vec{v})$ is the dipole element of the bound-continuum
transition.

The release time $t_{r}^{(i)}$ of trajectory $i$ is determined by the
saddle-point equation, 
\begin{equation}
\left. \frac{\partial S(t^{\prime })}{\partial t^{\prime }}\right\vert
_{t^{\prime }=t_{r}^{(i)}}=\frac{\left[ \vec{k}+\vec{A}(t_{r}^{(i)})\right]
^{2}}{2}+I_{p}=0,  \label{eq:12}
\end{equation}%
yielding complex values since $I_{p}>0$. The condition for different
trajectories to interfere is to reach the same final momentum $\vec{k}$ to
satisfy Eq. (\ref{eq:12}) with release times $t_{r}^{(i)}~(i=1,2,...,M).$
Whereas the interference condition involves the vector potential $\vec{A}$,
the electron trajectory is governed by the electrical field $\vec{F}$. We
now consider a periodic laser linearly polarized along the $z$ axis whose
laser field is $\vec{F}(t)=F_{0}\hat{z}\sin (\omega t)$, where $F_{0}$ is the
field amplitude. Accordingly, the vector potential is given by $\vec{A}(t)=%
\frac{F_{0}}{\omega }\hat{z}\cos (\omega t).$ There are two solutions of Eq.
(\ref{eq:12}) per optical cycle. The first solution in the $j$-th cycle is
given by%
\begin{equation}
t_{r}^{(j,1)}=\frac{2\pi (j-1)}{\omega }+\frac{1}{\omega }\cos ^{-1}\left[ -%
\tilde{\kappa}\right] ,  \label{tj1}
\end{equation}%
where $\tilde{\kappa}$ denotes the complex final momentum defined by 
\begin{equation}
\tilde{\kappa}=\kappa _{z}+i\sqrt{\gamma ^{2}+\kappa _{\bot}^{2}}
\label{kappatilde}
\end{equation}%
and $\kappa _{z}$ and $\kappa _{\bot}$ are the respective longitudinal and transversal 
components of the dimensionless scaled final momentum of the electron 
$\vec{\kappa}=\omega \vec{k}/F_{0}$. In Eq. (\ref{kappatilde}) 
$\gamma =\sqrt{2I_{p}}\,\omega /F_{0}$ is the Keldysh parameter.
The second solution fulfills 
\begin{equation}
t_{r}^{(j,2)}=\left\{ 
\begin{array}{ccc}
\frac{4\pi }{\omega }(j-\frac{1}{2})-t_{r}^{(j,1)} & \mathrm{if} & \kappa
_{z}\geq 0 \\ 
\frac{4\pi }{\omega }(j-1)-t_{r}^{(j,1)} & \mathrm{if} & \kappa _{z}<0.%
\end{array}%
\right.   \label{tj2}
\end{equation}%
In equations (\ref{tj1}) and (\ref{tj2}), $t_{r}^{(j,\alpha )}$ with $%
\alpha =1(2)$ denotes the early (late) release times within the $j$-th cycle.

For a given value of $\vec{k}$, the field strength for ionization at $%
t_{r}^{(j,\alpha )}$ is independent of $j$ and $\alpha $, then $\left\vert
F\left( t_{r}^{(j,\alpha )}\right) \right\vert =F_{0}\left\vert \sqrt{1-%
\tilde{\kappa}^{2}}\right\vert $. The ionization rate $\Gamma (\vec{k}%
)=|G(t_{r}^{(j,\alpha )},\vec{k})|^{2}e^{-2\Im \lbrack S(t_{r}^{(j,\alpha
)})]}$ is identical for all subsequent ionization bursts (or trajectories)
and, therefore, only a function of the time-independent final momentum $\vec{%
k}$ provided the ground-state depletion is negligible. As there are two
interfering trajectories per cycle, the total number of interfering
trajectories with final momentum $\vec{k}$ is $M=2N$, with $N$ being the
number of cycles involved in the laser pulse. Hence, the sum over
interfering trajectories [Eq.\ (\ref{bsol})] can be decomposed into those
associated with two release times within the same cycle and those associated
with release times in different cycles \cite{Arbo12,Arbo10b,Arbo10a}.
Consequently, the momentum distribution [Eq. (\ref{P})] can be written
within the SPA as 
\begin{equation}
\frac{dP}{d\vec{k}}=\Gamma (\vec{k})\left\vert \sum_{j=1}^{N}\,\sum_{\alpha
=1}^{2}e^{i\Re \lbrack S(t_{r}^{(j,\alpha )})]}\right\vert ^{2},
\label{interf}
\end{equation}%
where the second factor on the right hand side of Eq. (\ref{interf})
describes the interference of $2N$ trajectories with final momentum $\vec{k}$%
, where $t_{r}^{(j,\alpha )}$ is a function of $\vec{k}$ through equations (%
\ref{tj1}) and (\ref{tj2}).

The semiclassical action along one electron trajectory with release time $%
t_{r}^{(j,\alpha )}$ can be calculated within the SPA from Eq. (\ref{action}%
) up to a constant,%
\begin{equation}
S(t_{r}^{(j,\alpha )})=2U_{p}\left[ \left( \left\vert \tilde{\kappa}%
\right\vert ^{2}+\frac{1}{2}\right) t_{r}^{(j,\alpha )}+\frac{\sin (2\omega
t_{r}^{(j,\alpha )})}{4\omega }+2\frac{\kappa _{z}}{\omega }\sin (\omega
t_{r}^{(j,\alpha )})\right] ,  \label{sintra}
\end{equation}%
where the ponderomotive energy is given by $U_{p}=F_{0}^{2}/4\omega ^{2}$,
and $\left\vert \tilde{\kappa}\right\vert ^{2}=\left\vert \vec{\kappa}%
\right\vert ^{2}+\gamma ^{2}$ [see Eq. (\ref{kappatilde})]. The sum in Eq. (%
\ref{interf}) can be written as 
\begin{equation}
\sum_{j=1}^{N}\,\sum_{\alpha =1}^{2}e^{iS(t_{r}^{(j,\alpha
)})}=2\sum_{j=1}^{N}e^{i\bar{S}_{j}}\cos \left( \frac{\Delta S_{j}}{2}%
\right) ,\,  \label{sum}
\end{equation}%
where $\bar{S}_{j}=$ $\Re \left[ S(t_{r}^{(j,1)})+S(t_{r}^{(j,2)})\right] /2$
is the average action of the two trajectories released in cycle $j,$ and $%
\Delta S_{j}=\Re \left[ S(t_{r}^{(j,1)})-S(t_{r}^{(j,2)})\right] $ is the
accumulated action between the two release times $t_{r}^{(j,1)}$ and $%
t_{r}^{(j,2)}$ within the same $j$-th cycle. The average action depends
linearly on the cycle number $j$, so $\bar{S}_{j}=S_{0}+j\tilde{S}$,  
where $S_{0}$ is a constant which will drop out when the absolute value of
Eq. (\ref{sum}) is taken, and 
\begin{equation}
\tilde{S}=\left( 2\pi /\omega \right) \left( E+U_{p}+I_{p}\right) .
\label{Stilde}
\end{equation}%
In turn, due to discrete translation invariance in the time domain ($%
t\rightarrow t+2j\pi /\omega $), the difference of the action $\Delta S_{j}$
is independent of the cycle number $j$, which can be expressed (dropping the
subscript $j$) as%
\begin{eqnarray}
\Delta S &=&\frac{-2U_{p}}{\omega } \: \Re \left[ \left( 1+2|\tilde{\kappa}%
|^{2}\right) \mathrm{sgn}(\kappa _{z})\cos ^{-1}(\mathrm{sgn}(\kappa _{z})\ 
\tilde{\kappa})\right.   \label{DS} \\
&&\left. -\left( 4\kappa _{z}-\tilde{\kappa}\right) \sqrt{1-\tilde{\kappa}%
^{2}}\right] ,  \notag
\end{eqnarray}%
where $\mathrm{sgn}$ denotes the sign function that accounts for positive
and negative longitudinal momentum $k_{z}$.

After some algebra, Eq. (\ref{interf}) can be rewritten as an equation of a
diffraction grating of the form \cite{Arbo12,Arbo10b,Arbo10a}, 
\begin{equation}
\frac{dP}{d\vec{k}}=4\,\Gamma (\vec{k})\underbrace{\cos ^{2}\left( \frac{%
\Delta S}{2}\right) }_{F(\vec{k})}\underbrace{\left[ \frac{\sin \left( N%
\tilde{S}/2\right) }{\sin \left( \tilde{S}/2\right) }\right] ^{2}}%
_{B(k)}\quad ,  \label{eq:24}
\end{equation}%
where the interference pattern can be factorized into two contributions: (i)
the interference stemming from a pair of trajectories within the same cycle
(intracycle interference), governed by $F(\vec{k}),$ and (ii) the
interference stemming from trajectories released at different cycles
(intercycle interference) resulting in the well-known ATI peaks given by $%
B(k)$ (see Ref. \cite{Faisal05}). The intracycle interference arises from
the superposition of pairs of trajectories separated by a time slit $\Delta
t=t_{r}^{(j,1)}-t_{r}^{(j,2)}$ of the order of less than half a period of
the laser pulse, i.e., $\Re(\Delta t)<\pi /\omega $, while the difference
between $t_{r}^{(j,\alpha )}$ and $t_{r}^{(j+1,\alpha )}$ is $2\pi /\omega $%
, i.e., the optical period of the laser. It is worth to note that whereas
the intracycle factor $F(\vec{k})$ depends on the angle of emission, the
intercycle factor $B(k)$ depends only on the absolute value of the final
momentum (or energy). Eq.\ (\ref{eq:24}) may be viewed as a diffraction
grating in the time domain consisting of $N$ slits with an interference
factor $B(k)$ and diffraction factor $F(\vec{k})$ for each slit.
In the following subsection we make use of the Fourier theory of last subsection
to analyze the moir\'{e} patterns in the doubly differential momentum 
distribution [Eq. (\ref{eq:24})].

\subsection{Formation of moir\'{e} patterns from inter- and intracycle
interference}

The intercycle principal maxima fulfill the equation $\tilde{S}=2n\pi ,$
leading to the ATI energies $E_{n}=n\omega -$ $U_{p}-I_{p}$ in agreement
with the conservation of energy for the absorption of $n$ photons.
Therefore, in the doubly differential momentum distribution, the 2D
intercycle grid follows the relation between the parallel and perpendicular
momenta $k_{\perp }^{\mathrm{inter}}(n)=\sqrt{2(n\omega -I_{p}-U_{p})-\left[
k_{z}^{\mathrm{inter}}(n)\right] ^{2}}.$ The spacing between two consecutive
maxima can be easily calculated for $E_{n}=k_{z}^{2}/2,$ (provided $k_{\perp
}=0$) as%
\begin{eqnarray}
\frac{\lbrack k_{z}^{\mathrm{inter}}(n+1)]^{2}-[k_{z}^{\mathrm{inter}%
}(n)]^{2}}{2} &\simeq &k_{z}\Delta k_{z}^{\mathrm{inter}}  \notag \\
&\Rightarrow &\Delta k_{z}^{\mathrm{inter}}\simeq \frac{\omega }{k_{z}}=%
\frac{1}{\alpha \ \kappa _{z}},  \label{Dkzinter}
\end{eqnarray}%
where $\alpha =4U_{p}/F_{0}=F_{0}/\omega ^{2}$ is the quiver amplitude of
the escaping electron, $\vec{\kappa}=(\omega /F_{0})\vec{k}$; and in the
last line we have used that $E_{n+1}-E_{n}=\omega .$

The intracycle maxima correspond to the equation $\Delta S=2m\pi $ with
integer $m.$ Equivalently to the intercycle case, the intracycle spacing can
be calculated as 
\begin{eqnarray}
\frac{\Delta S(k_{z}+\Delta k_{z})}{2}-\frac{\Delta S(k_{z})}{2} &\simeq &%
\frac{1}{2}\left. \frac{\partial \Delta S(k_{z})}{\partial k_{z}}\right\vert
_{k_{\perp }=0}\Delta k_{z}^{\mathrm{intra}}  \label{Dkzintrap} \notag \\
&\Rightarrow &\Delta k_{z}^{\mathrm{intra}}\simeq \frac{2\pi }{\left\vert 
\frac{\partial \Delta S(k_{z})}{\partial k_{z}}\right\vert _{k_{\perp }=0}}.
\end{eqnarray}%
After a bit of algebra, the derivative of the accumulated action with
respect to the parallel momentum can be written in a close form and, thus,
the intracycle spacing reads%
\begin{equation}
\Delta k_{z}^{\mathrm{intra}}=\frac{\pi }{\alpha \left\vert \Re \left[
\kappa _{z}\cos ^{-1}(\kappa _{z}+i\gamma )-\sqrt{1-(\kappa _{z}+i\gamma
)^{2}}\right] \right\vert }.  \label{Dkzintra}
\end{equation}%
In Eq. (\ref{Dkzintra}) we have considered forward emission, i.e., $%
k_{z}\geq 0.$ We have an analogous result for backward emission.

According to Eq. (\ref{eq:24}), the transformations from the 1D grating to
the inter- and intracycle 2D grating are $T_{1}(k_{z},k_{\perp })=\tilde{S}%
/2 $ given by Eq. (\ref{Stilde}) and $T_{2}(k_{z},k_{\perp })=\Delta S/2$
given by Eq. (\ref{DS}). Therefore, we can write the $(k_{1},k_{2})-$order
compound transformation $T(k_{z},k_{\perp })=k_{1}T_{1}(k_{z},k_{\perp
})+k_{2}T_{2}(k_{z},k_{\perp })$ as%
\begin{equation}
T(k_{z},k_{\perp })=k_{1}\frac{\tilde{S}}{2}+k_{2}\frac{\Delta S}{2}.
\label{transformation}
\end{equation}

By eye inspection [at least for the lowest orders $%
(k_{1},k_{2})=(1,-1),(2,-1)$, and $(1,-2)$] the function $T(k_{z},k_{\perp
}) $ exhibits one global minimum for forward (and backward) emission,
which corresponds to the center of the side ring.
The minimum can be easily found as%
\begin{equation}
\vec{\nabla}T(k_{z},k_{\perp })=\left( \frac{\partial T(k_{z},k_{\perp })}{%
\partial k_{z}},\frac{\partial T(k_{z},k_{\perp })}{\partial k_{\perp }}%
\right) =0.  \label{gradT}
\end{equation}

One find that $k_{\perp }=0$ is solution of $\partial \tilde{S}/\partial
k_{\perp }=0,$ and $\partial \Delta S/\partial k_{\perp }=0,$ separately and
independently of the value of $k_{z}.$ Therefore, $k_{\perp }=0$ is solution
of the second component of Eq. (\ref{gradT}), $\partial T(k_{z},k_{\perp
})/\partial k_{\perp }=(k_{1}/2)(\partial \tilde{S}/\partial k_{\perp
})+(k_{2}/2)(\partial \Delta S/\partial k_{\perp })=0,$ irrespective of the
values of $k_{1}$, $k_{2}$, and $k_{z}.$ This means that the center of the
moir\'{e} rings lay along the $k_{z}$ axis ($k_{\perp }=0$).

Now, with the restriction $k_{\perp }=0,$we formally solve the first
component of Eq. (\ref{gradT})%
\begin{equation}
\frac{\partial T(k_{z},k_{\perp })}{\partial k_{z}}=\frac{k_{1}}{2}\frac{%
\partial \tilde{S}}{\partial k_{z}}+\frac{k_{2}}{2}\frac{\partial \Delta S}{%
\partial k_{z}}=0.  \label{dTdkz}
\end{equation}%
The derivative in the first term of right hand side of Eq. (\ref{dTdkz}) can be easily written
as $\partial \tilde{S}/\partial k_{z}=2\pi k_{z}/\omega =2\pi /\Delta k_{z}^{%
\mathrm{inter}},$ where we have used equations (\ref{Stilde}) and (\ref%
{Dkzinter}). Doing the same with the derivative in the second term of Eq. (%
\ref{dTdkz}), we get that $\partial \Delta S/\partial k_{z}=2\pi /\Delta
k_{z}^{\mathrm{intra}}.$ Therefore, Eq. (\ref{dTdkz}) can be written as%
\begin{equation}
\frac{\partial T(k_{z},k_{\perp })}{\partial k_{z}}=\pi \left( \frac{k_{1}}{%
\Delta k_{z}^{\mathrm{inter}}}+\frac{k_{2}}{\Delta k_{z}^{\mathrm{intra}}}%
\right) =0,
\end{equation}%
which is equivalent to%
\begin{equation}
k_{1}\Delta k_{z}^{\mathrm{intra}}=-k_{2}\Delta k_{z}^{\mathrm{inter}}.
\label{spacings}
\end{equation}%
The principal moir\'{e} rings is given by the lowest order $%
(k_{1},k_{2})=(1,-1),$ which means that the intra- and intercycle spacings
should be the same, i.e., $\Delta k_{z}^{\mathrm{intra}}=\Delta
k_{z}^{\mathrm{inter}}$. This result provides the position of the center of the
principal moir\'{e} pattern. Higher order moir\'{e} patterns, i.e., $%
(k_{1},k_{2})=(2,-1)$ and $(1,-2)$, denote the secondary moir\'{e} rings
whose centers are positioned along the $k_{z}$ axis at the $k_{z}$ value
which makes the intercycle spacing the double of the intracycle one, i.e., $%
2\Delta k_{z}^{\mathrm{intra}}=\Delta k_{z}^{\mathrm{inter}},$ and the
intracycle spacing the double of the intercycle one, i.e., $\Delta k_{z}^{%
\mathrm{intra}}=2\Delta k_{z}^{\mathrm{inter}}$, respectively.

\section{Results and discussion}

\label{results} 

\begin{figure}[tbp]
\centering
\includegraphics[width=0.65\textwidth]{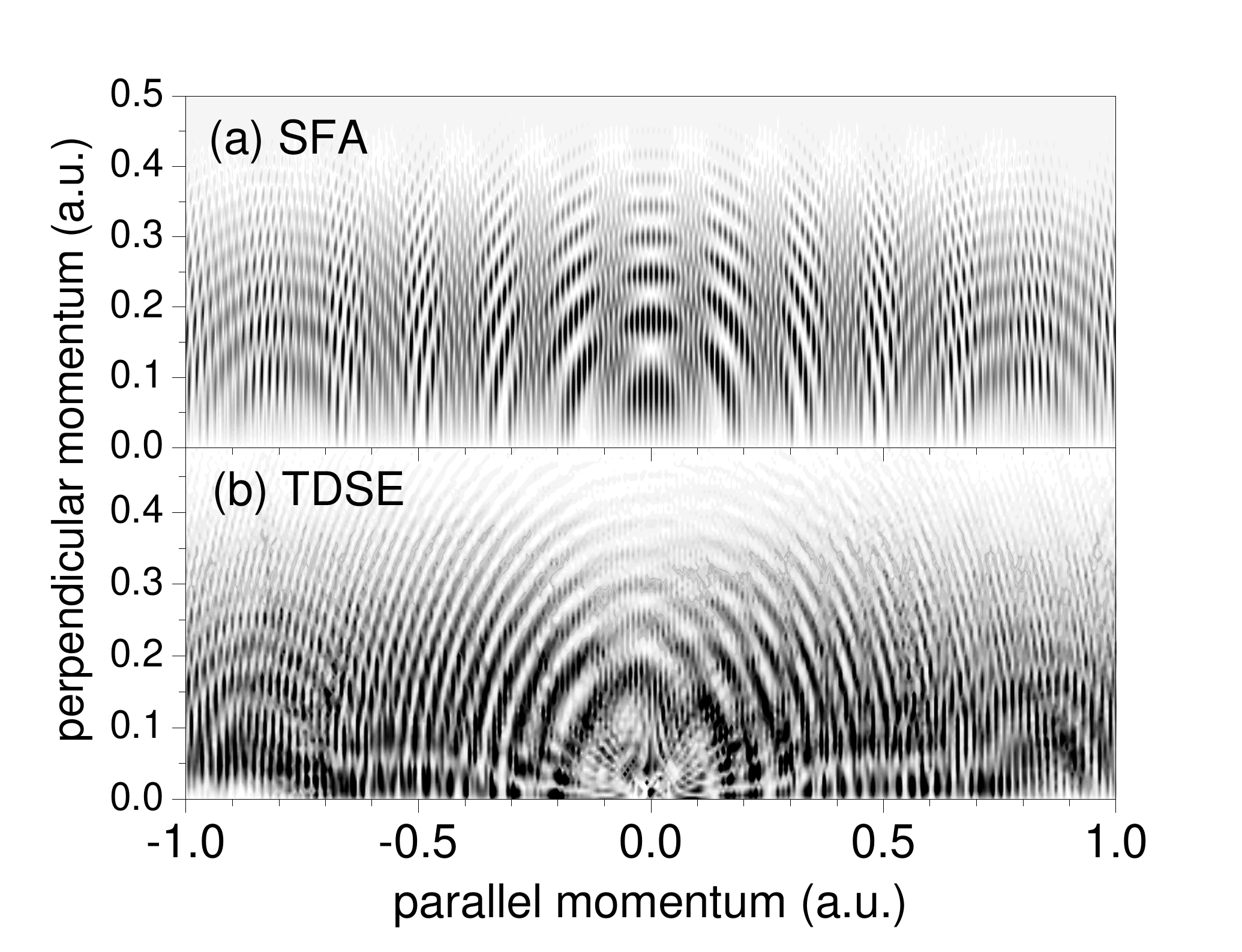}
\caption{Momentum distributions (linear grey scale) after
interaction of a midinfrared laser pulse with a hydrogen atom. (a) SFA and
(b) TDSE \cite{Lemell13,tongprivcom}. The cosine-like pulse has a peak field $%
F_{0}=0.0533 $ ($I=10^{14}$ W/cm$^{2}$), frequency $\omega =0.01424$ ($%
\lambda =3200$ nm) and a sin$^{2}$ envelope with total pulse duration of
eight cycles.
}
\label{q-distributions}
\end{figure}

In Fig. \ref{q-distributions} (a) and (b) we show the doubly differential electron momentum
distribution within the SFA [equations (\ref{P}) and (\ref{Tif})], and TDSE 
\cite{Lemell13,tongprivcom} after ionization of atomic hydrogen by an intense ($I=10^{14}
$ W/cm$^{2}$) midinfrared ($\lambda =3200$ nm or equivalently $\omega
=0.001424$ a.u.) sine- pulse of eight-cycle of total duration with a $\sin
^{2}$ envelope. The intercycle pattern appear as concentric (ATI) rings
centered at threshold. In the TDSE momentum distribution, the characteristic
bouquet-shape structure due to interference of electron trajectories
oscillating about the Kepler trajectory is clearly observed \cite%
{Arbo06a,Arbo08b}. The bouquet-shape structure is absent in the SFA since it lacks of
the effect of the Coulomb potential on the escaping trajectories.
At both sides of the ATI rings, two symmetrical annular structures at $%
|k_{z}|\simeq 0.82$ are observed in both (SFA and TDSE) approaches. As far
as we know, these side rings has not been studied. As the SFA does not
consider rescattering electrons, we must discard this effect as a possible
explanation for the formation of the side rings. In the rest of the paper we
identify the origin of these rings with the aid of the semiclassical model
and the theory of moir\'{e} patterns.

\begin{figure}[tbp]
\centering
\includegraphics[width=0.65\textwidth]{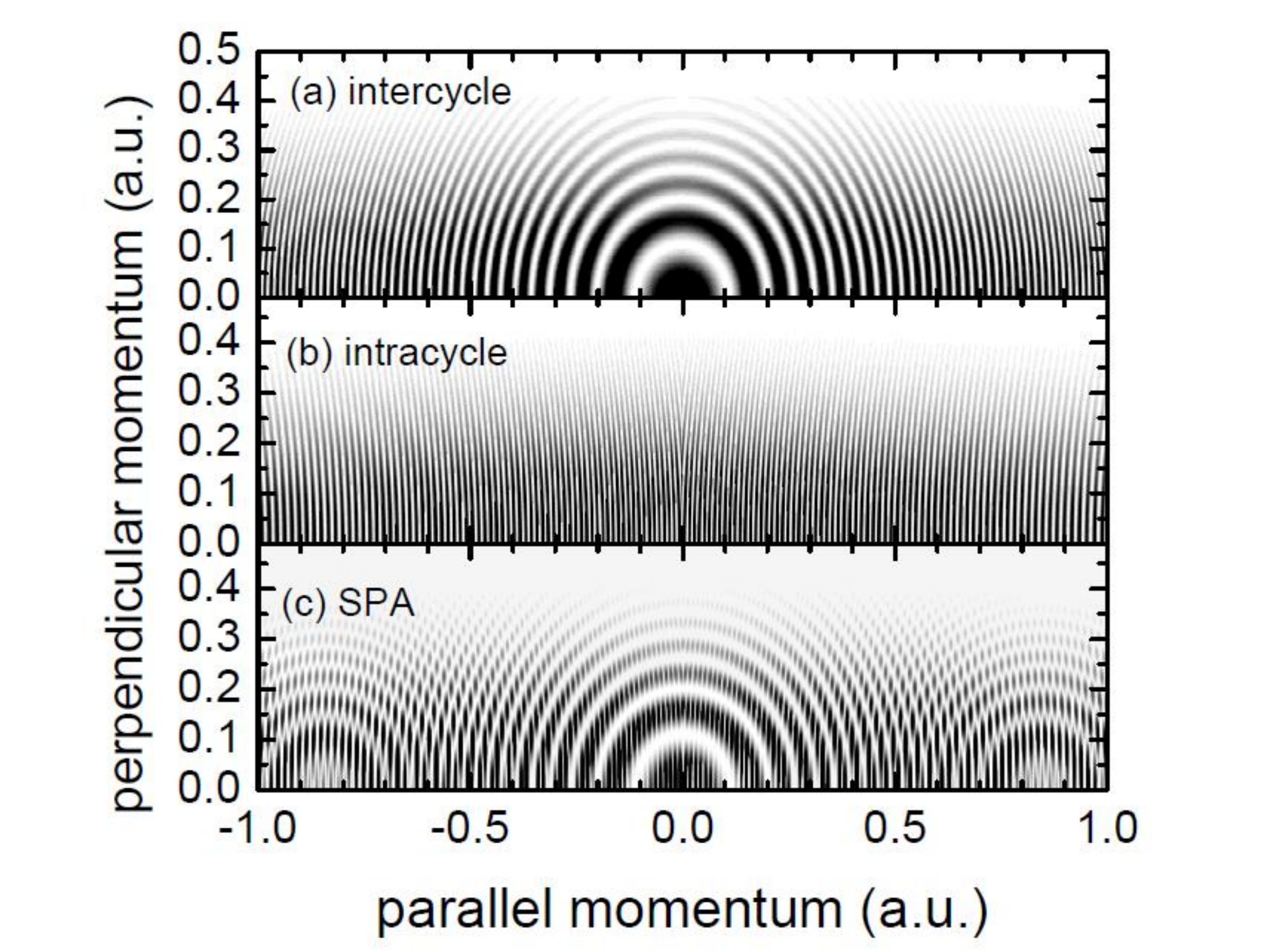}
\caption{SPA doubly differential momentum distribution (linear grey
scale) of Eq. (\ref{eq:24}). (a) Intracycle interference: $4\Gamma (\vec{k}%
)F(\vec{k}),$ (b) intercycle interference: $4\Gamma (\vec{k})B(k)$ for $N=2$
cycles, and (c) total (intra- and intercycle) interference $4\Gamma (\vec{k}%
)F(\vec{k})B(k)$ for $N=2$ cycles. The laser parameters are $F_{0}=0.0533$ ($%
I=10^{14}$ W/cm$^{2}$) and frequency $\omega =0.01424$ ($\lambda =3200$ nm).
}
\label{scm-distributions}
\end{figure}

The interplay between the intercycle interference [factor $B(k)$ in Eq. (\ref{eq:24})]
and the intracycle interference [factor $F(\vec{k})$ in Eq. (\ref{eq:24})]
controls the doubly differential distribution of direct ATI electrons for lasers
\cite{Arbo10a,Arbo10b,Arbo12}. Firstly, we examine the
intercycle interference within the SPA by setting the intracycle factor to be $F(\vec{k})=1$
and $N=2$ in Eq. (\ref{eq:24}) for the same laser parameters as in Fig. 
\ref{q-distributions}, except the duration and envelope, we use $N=2$ cycles of duration.
The factor $B(k)$ reduces to the two-slit Young interference expression 
$B(k)=4\cos ^{2}\left[ \pi /\omega \left( \tilde{S}/2 \right) \right]$, 
where $\tilde{S}$ is given by Eq. (\ref{Stilde}). We
plot the corresponding SPA doubly differential momentum distribution in Fig.
2 (a), where we can observe concentric rings with radii $k_{n}=\sqrt{%
2E_{n}}$. The intracycle interference arises from the superposition of two
trajectories released within the same optical cycle, i.e., $\alpha =1,2$ and 
$N=1$ in Eq. (\ref{eq:24}) or, equivalently, $4\,\Gamma (\vec{k})F(\vec{k}),
$ since $B(k)=1$ in this case. In Fig. \ref{scm-distributions} (b), we see that the SPA intracycle
interference pattern gives approximately vertical thin stripes which bend to
the higher energy region as the transverse momentum grows. The width of the
stripes increases with the energy. In order to analyze the complete pattern
stemming from all four interfering trajectories in the two-cycle pulse, the
composition of the intercycle and intracycle interference patterns of Figs.
2 (a) and (b) gives the SPA momentum distribution of Fig. \ref{scm-distributions} (c).
We can see
that a grosser structure emerges as two side rings centered at $k_{z}\simeq
\pm 0.83$ and $k_{\perp }\simeq 0$, and two less visible rings centered at $%
k_{z}\simeq \pm 0.5$ and $k_{\perp }\simeq 0$. If we consider longer pulses,
the contrast of intercycle factor $B(k)$ will increase as $N$ increases. For
example, the ATI rings will become narrower and $N-2$ secondary rings will
appear between two consecutive principal ATI rings. On the other side, the
intracycle factor $F(\vec{k})$ is independent of the number of cycles $N$
involved in the laser pulse and, in consequence, the intracycle interference
pattern remains unchanged. This is strictly valid provided we consider a
flattop pulse in the SPA. Moreover, we have checked that the position of the
side rings is independent of the pulse duration (not shown).

\begin{figure}[tbp]
\centering
\includegraphics[angle = 0, trim = 1mm 4mm 1mm 5mm, clip, width=0.8\textwidth]{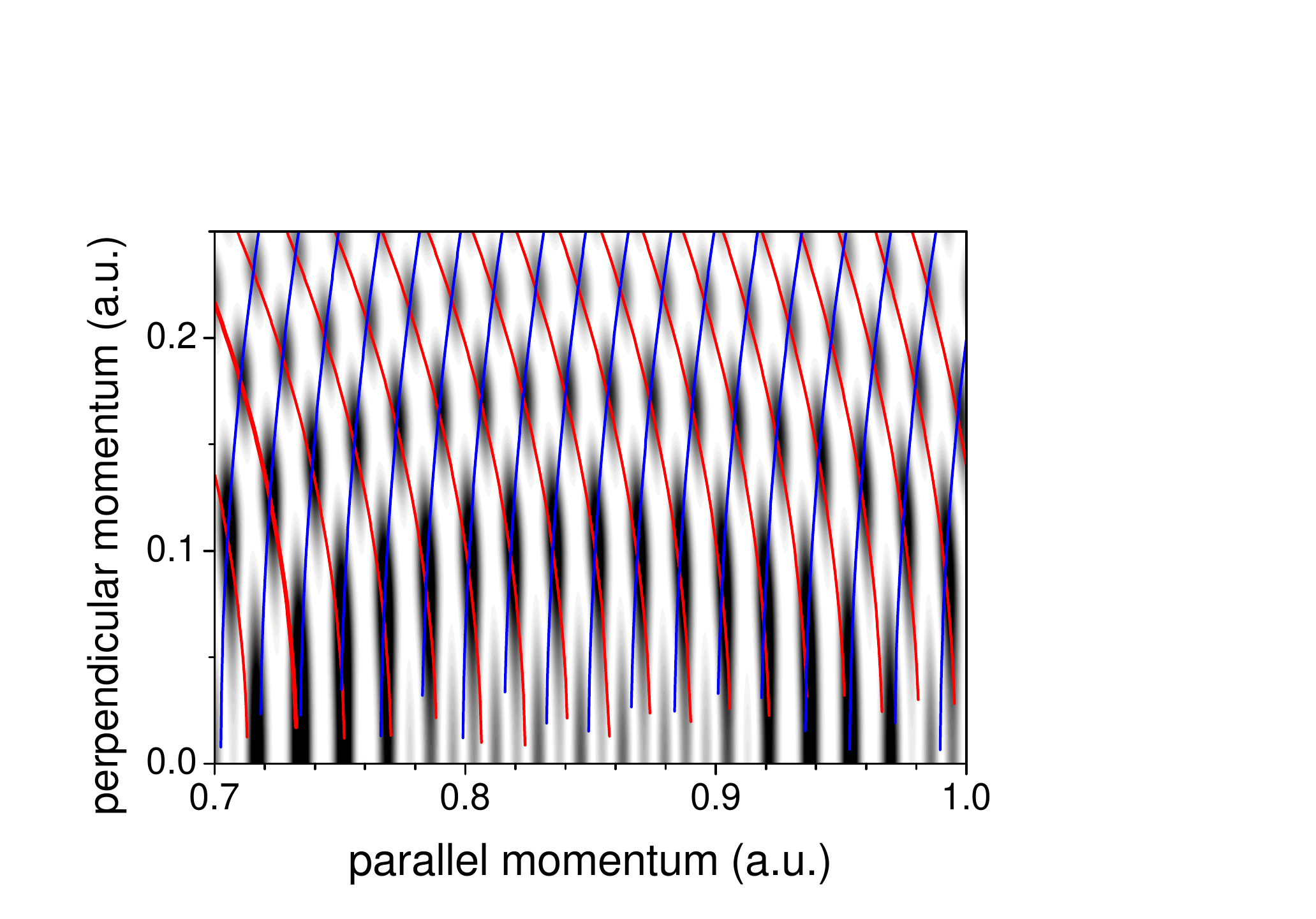}
\caption{Magnification of SPA doubly differential momentum
distribution in Fig. \ref{scm-distributions} (c) (linear gray scale). On top of it we have drawn
the different intercycle maxima, i.e., $\tilde{S}=2n\pi $ (with integer $n$)
in blue and the different intracycle interference maxima, i.e., $\Delta
S=2m\pi $ (with integer $m$) in red. The local maxima of the doubly
differential momentum distribution coincide with the intersection of the
inter- and intracycle maxima.
}
\label{grid}
\end{figure}

In the Fig. \ref{grid}, we show in red the maxima of the intercycle interference
pattern, i.e., $\tilde{S}=2n\pi $ with $n$ integer, given by the
conservation of energy relation and in blue the maxima of the intracycle
interference pattern, i.e., $\Delta S=2m\pi $ with $m$ integer, on top of
the SPA\ doubly differential momentum distribution of Fig. \ref{scm-distributions} (c) in the
region of the main side ring in the forward direction. We clearly see how
the intersections of the inter- and intracycle grids coincide with the
different local maxima of the distribution forming an annular structure.
Contrarily to the intercycle grid, the intracycle grid does not have an
explicit form and must be solved numerically.

\begin{figure}[tbp]
\centering
\includegraphics[width=0.8\textwidth]{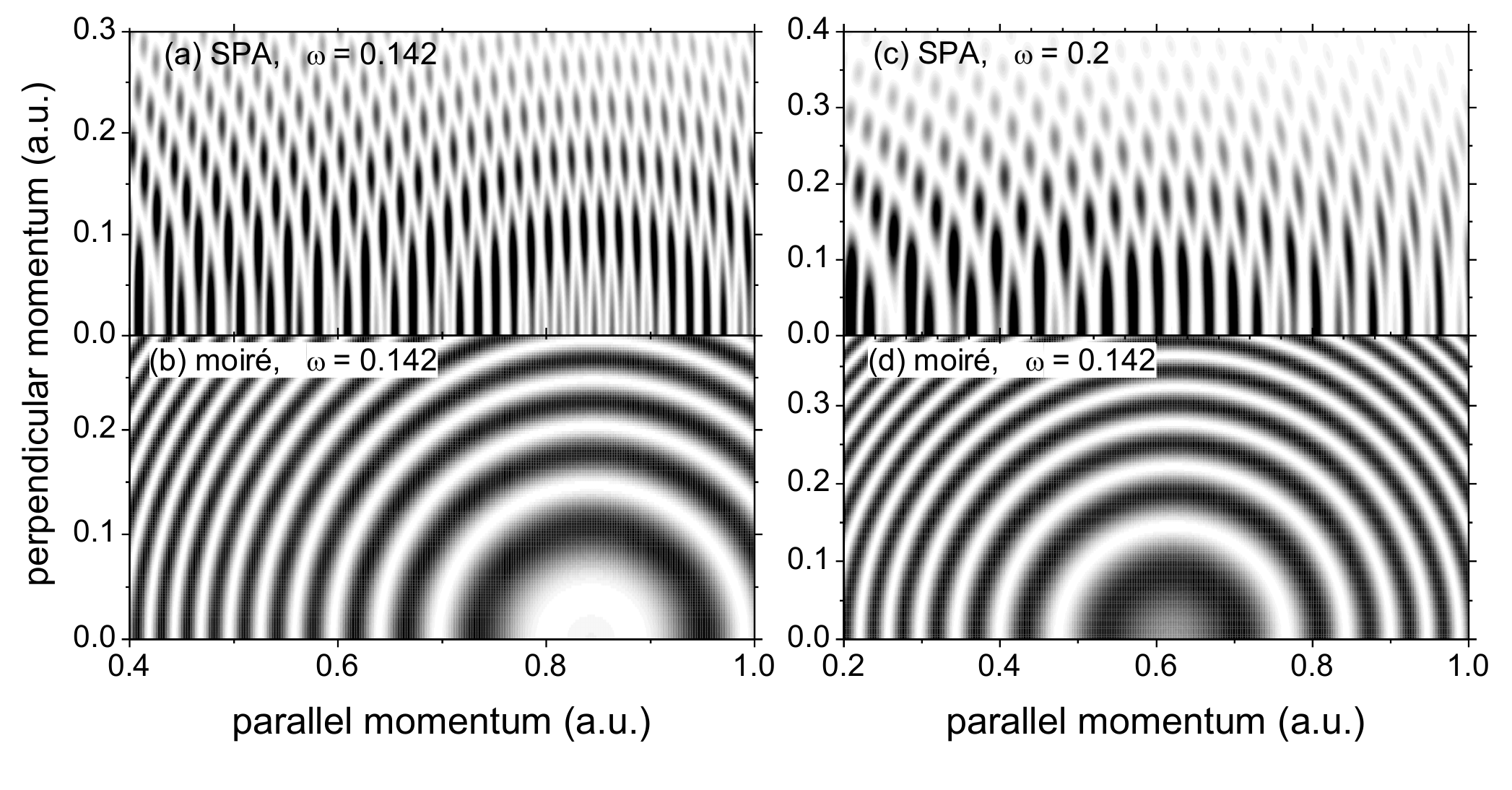}
\caption{SPA doubly differential momentum distribution (linear grey
scale) of Eq. (\ref{eq:24}) [(a and (c)] and the corresponding (1,-1) moir%
\'{e} pattern [(b) and (d)]. The laser frequency is $\omega =0.1424$ for (a)
and (b) and $\omega =0.2$ for (c) and (d). The rest of the laser parameters
are the same as in Fig. \ref{scm-distributions} and Fig. \ref{grid}.
}
\label{principal}
\end{figure}

In Fig. \ref{principal} (a) we show a closeup of the SPA doubly differential momentum distribution for
the same laser parameters as in figures \ref{scm-distributions} and \ref{grid}.
The side ring centered at $(k_{z},k_{\perp })\simeq (0.83,0)$ is clearly seen.
In Fig. \ref{grid} (b) we plot the principal moir\'{e} ring, 
i.e., $\cos ^{2}[(T(k_{z},k_{\perp })]$ ,
where the transformation $T(k_{z},k_{\perp })$ is given by Eq. (\ref%
{transformation}) for $(k_{1},k_{2})=(1,-1)$. We see that the shape and
position of the moir\'{e} pattern in Fig. \ref{principal} (b) coincide with the side ring
of the doubly differential momentum distribution in Fig. \ref{principal} (a). When the
laser frequency is increased to $\omega =0.2$, the principal side ring
shifts horizontally towards less energetic domains and is centered at $%
(k_{z},k_{\perp })\simeq (0.62,0),$ as can be observed in Fig. \ref{principal} (c). The corresponding
moir\'{e} pattern in Fig. \ref{principal} (d), also shifts accurately reproducing the side
ring. This is a confirmation of the application of the theory of the principal moir\'{e}
patterns posed in the last section to atomic ionization in the midinfrared range.

\begin{figure}[tbp]
\centering
\includegraphics[width=1\textwidth]{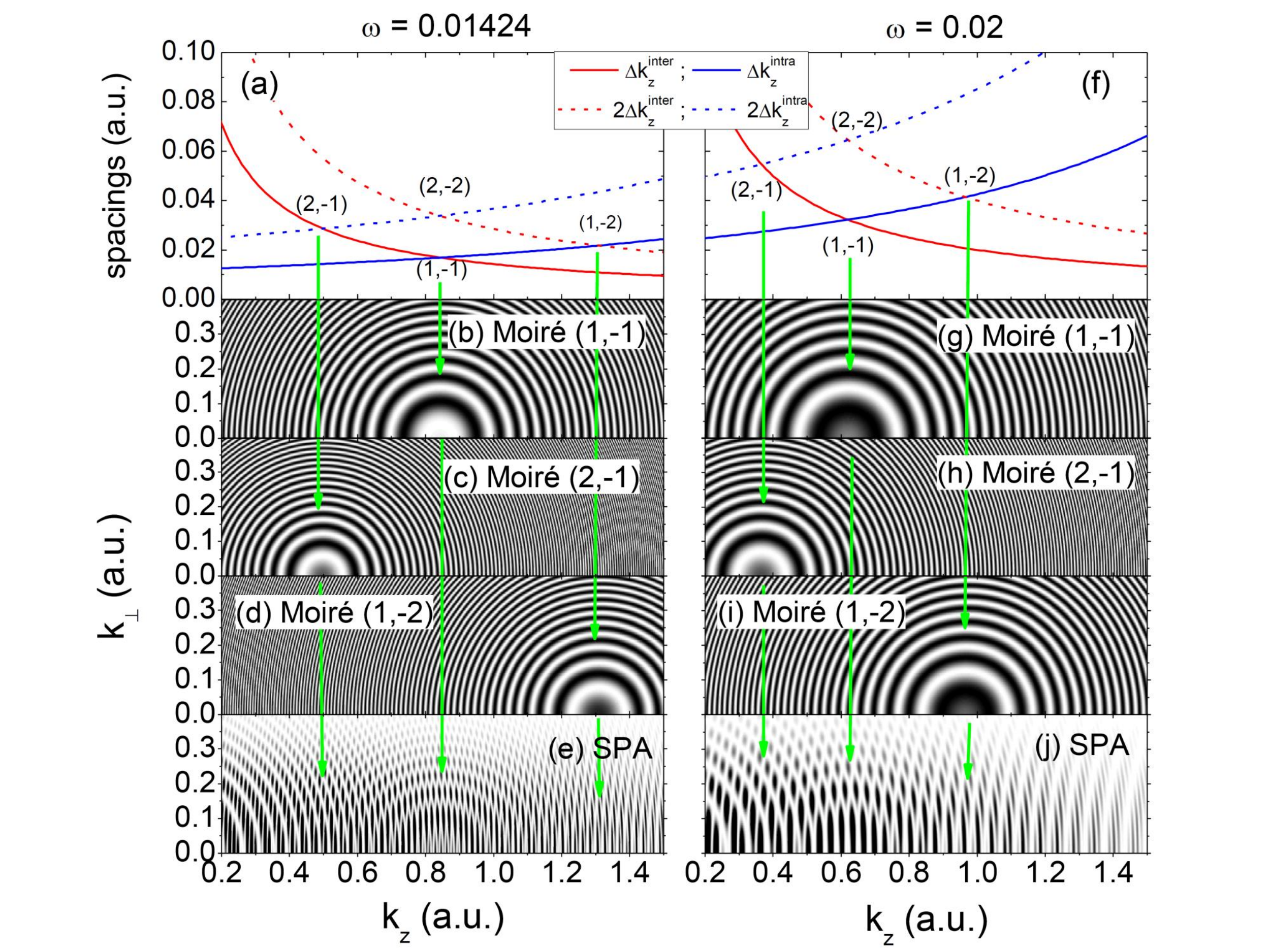}
\caption{Inter- and intracycle spacings with their first harmonic
(twice the spacings) of Eq. (\ref{Dkzinter}) and Eq. (\ref{Dkzintra}),
respectively [(a) and (f)]. Main moir\'{e} (1,-1) pattern [(b) and (g)] and
secondary moir\'{e} (1,-2) [in (c) and (h)] and (2,-1) [in (d) and (i)]
patterns. SPA doubly differential momentum distribution (linear grey scale)
of Eq. (\ref{eq:24}) in (e) and (j). For the first column (a-e) $\omega
=0.1424$ and for the second column (f-j) $\omega =0.2.$ The rest of the
laser parameters are the same as in figures 2, 3, and 4.
}
\label{secondary}
\end{figure}

To fully confirm the theory of the moir\'{e} patterns, we plot the inter-
and intracycle spacings for $\omega =0.01424$ in Fig. \ref{secondary} (a) and $\omega =0.2$
in Fig. \ref{secondary} (b) in solid line, together with the double of the corresponding
spacings in dash lines. The rest of the laser parameters are the same as in
previous figures. In the second row of Fig. \ref{secondary} the principal moir\'{e} ring
is centered at the $k_{z}$ value which corresponds the intersection of the
inter- and intracycle spacings in agreement with Eq. (\ref{spacings}) for $%
(k_{1},k_{2})=(1,-1),$ i.e., $\Delta k_{z}^{\mathrm{intra}}=\Delta k_{z}^{%
\mathrm{inter}}.$ We see that the center of the moir\'{e} pattern is
situated at $k_{z}=0.84$ for $\omega =0.01424$ in Fig. \ref{secondary} (b), whereas it is
at $k_{z}=0.62$ for $\omega =0.02$ in Fig. \ref{secondary} (g). In the third row we see
that the secondary moir\'{e} pattern of order $(2,-1)$ is centered at the
intersection of the intercycle spacing and twice the intracycle spacing in
agreement with Eq. (\ref{spacings}), i.e., $2\Delta k_{z}^{\mathrm{intra}%
}=\Delta k_{z}^{\mathrm{inter}}.$ The center of the $(2,-1)$ is at $%
k_{z}=0.5 $ for $\omega =0.01424$ in Fig. \ref{secondary} (c), whereas it is at $%
k_{z}=0.37 $ for $\omega =0.02$ in Fig. \ref{secondary} (h). In the fourth line we see
that the secondary moir\'{e} pattern of order $(1,-2)$ is centered at the
intersection of twice the intercycle spacing and the intracycle spacing in
agreement with Eq. (\ref{spacings}), i.e., $\Delta k_{z}^{\mathrm{intra}%
}=2\Delta k_{z}^{\mathrm{inter}}.$ The center of the $(1,-2)$ is at $%
k_{z}=0.1.31$ for $\omega =0.01424$ in Fig. \ref{secondary} (d), whereas it is at $%
k_{z}=0.97$ for $\omega =0.02$ in Fig. \ref{secondary} (i). For the sake of completeness,
in the last row, we show the complete doubly differential momentum
distribution within the SPA. We clearly observe how the principal $(1,-1)$
and secondary $(2,-1)$ and $(1,-2) $ moir\'{e} patterns are mirrored in the
momentum distribution. Not only does the center of the moir\'{e} rings
coincide with the prediction of Eq. (\ref{spacings}) and observed in the
corresponding moir\'{e} patterns $\cos ^{2}[(T(k_{z},k_{\perp })],$ but also
the radii of the moir\'{e} rings perfectly agrees with the momentum
distribution. We want to point out that not only are the positions of the 
center of the moir\'{e} structures described by the theory but also the radii of the rings
themselves are fully reproduced. No counterpart of the secondary moir\'{e}
rings $(2,-1)$ and $(1,-2)$ are observed in the SFA and TDSE doubly
differential momentum distribution of Fig. \ref{q-distributions} since their visibility 
is very poor. In conclusion, Fig. \ref{secondary} provides a fully confirmation
of the application of the theory of the moir\'{e} patterns for principal and 
secondary rings to the formation of the side rings in the ionization of atomic hydrogen
by midinfrared lasers.

\begin{figure}[tbp]
\centering
\includegraphics[width=.5\textwidth]{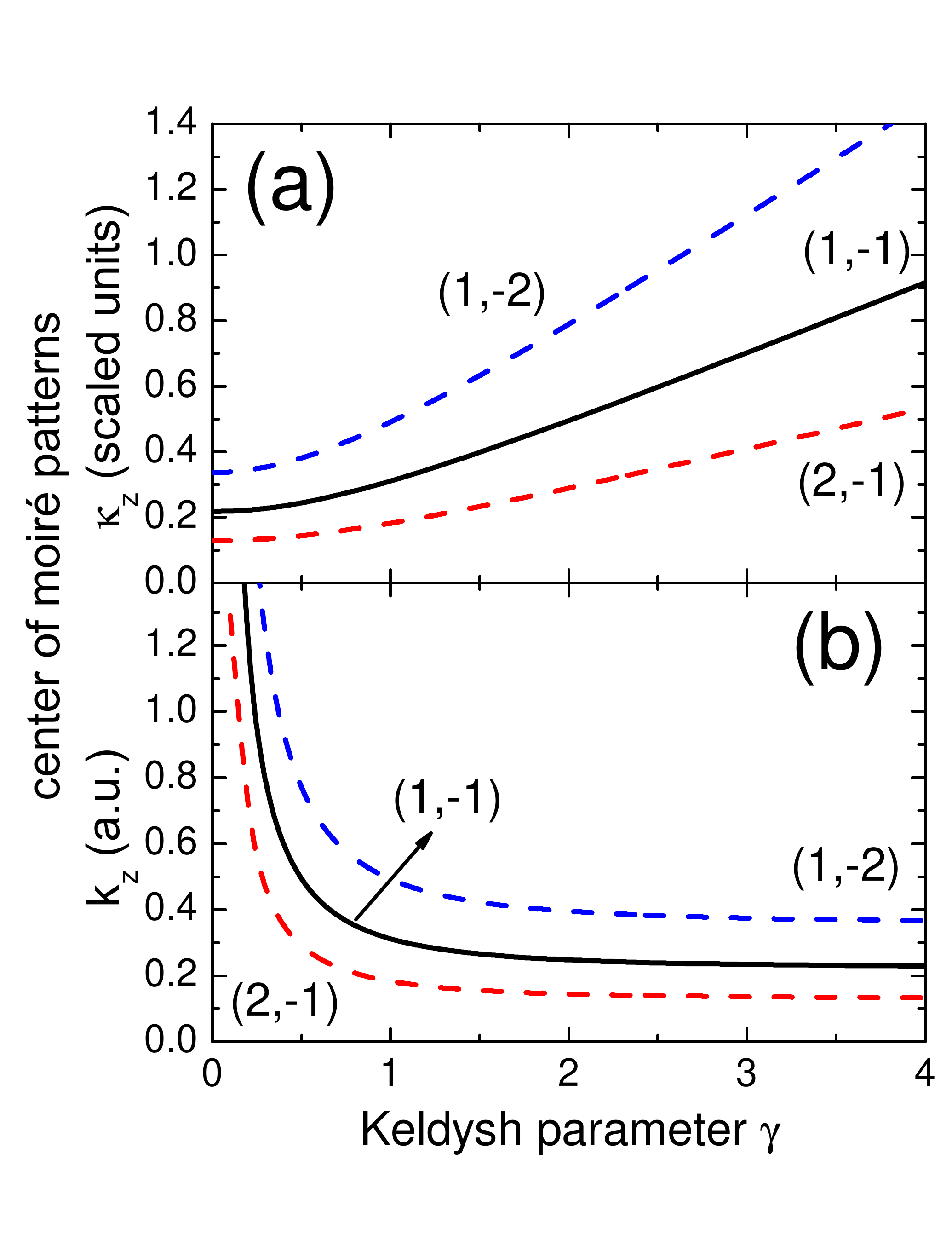}
\caption{Position of the center of the main and secondary moir\'{e}
patterns in units of the scaled parallel momentum $\kappa _{z}$ (a) and the
parallel momentum $k_{z}$ (b) as a function of the Keldysh parameter $\gamma $.
}
\label{centers}
\end{figure}

From equations (\ref{Dkzinter}) and (\ref{Dkzintra}) we see that the
positions of the centers of the principal and secondary rings in terms of $%
\kappa _{z}$ do not depend on the laser amplitude $F_{0}$ and frequency $%
\omega $ independently, but through the Keldysh parameter $\gamma .$ With
this in mind, in Fig. \ref{centers} we plot the center of the principal $(1,-1)$ and
secondary $(2,-1)$ and $(1,-2)$ moir\'{e} rings as a function of $\gamma .$
In Fig. \ref{centers} (a), we observe that the position of the center of both principal
and secondary rings measured in terms of the scaled $\kappa _{z}$ momentum
increases with the Keldysh parameter. In the tunneling limit ($\gamma \ll 1$%
), the position of the center of the moir\'{e} rings approach to constant
values $\kappa _{zc}^{(1,-1)}=0.217,$ $\kappa _{zc}^{(2,-1)}=0.128,$ and $%
\kappa _{zc}^{(1,-1)}=0.337.$ This result leads to a scale law for the
position of the moir\'{e} rings in the tunneling regime%
\begin{eqnarray}
k_{zc}^{(1,-1)} &=&0.217\frac{F_{0}}{\omega },  \notag \\
k_{zc}^{(2,-1)} &=&0.128\frac{F_{0}}{\omega },  \label{scale} \\
k_{zc}^{(1,-2)} &=&0.337\frac{F_{0}}{\omega },  \notag
\end{eqnarray}%
where we have used that $\vec{\kappa}=(\omega /F_{0})\vec{k}.$ This means
that the position of the center of the moir\'{e} rings scale as the inverse
of the Keldysh parameter $\gamma ^{-1},$ which is observed in Fig. \ref{centers} (b) in
the tunneling regime ($\gamma <1$). In the multiphoton regime ($\gamma >1$),
we see in Fig. \ref{centers} (a), that the center of the moir\'{e} rings follows an
approximate linear behavior with $\gamma ,$ i.e., $\kappa
_{zc}^{(1,-1)}\simeq 0.214\gamma ,$ $\kappa _{zc}^{(2,-1)}\simeq 0.124\gamma
,$ and $\kappa _{zc}^{(1,-1)}\simeq 0.345\gamma ,$ which is consistent with
the asymptotic values for the center of the moir\'{e} rings in the
multiphoton limit ($\gamma \gg 1$), observed in Fig. \ref{centers} (b). The
proportionality coefficients were calculated as the average slope of curves
in Fig. \ref{centers} (a) between $\gamma =3$ and $4$. The approximate agreement between
the asymptotic values for the center of the moir\'{e} fringes in the
multiphoton limit and the coefficients of Eq. (\ref{scale}) in the tunneling
regime is very suspicious to say that it is pure coincidence and deserves
more investigation.

\section{\label{conc}Conclusions}


We have presented a study of interference effects observed in the direct
ionization of atoms subject to multicycle laser pulses with wavelength in
the range of the midinfrared. In the framework of the SPA we describe the
full differential electron momentum distribution and identify side rings
calculated within the SFA and TDSE \cite{Lemell13} as the moir\'{e} fringes
due to the interplay between the intra- and intercycle interferences of
electron trajectories in photoelectron 3D momentum distribution. A whole
family of moir\'{e} fringes of varying visibility was characterized. An
analytical expression for the moir\'{e} patterns within a Fourier analysis
is presented showing an excellent agreement with the numerical calculations.
The principal (secondary) side rings are centered along the parallel momentum axis
($k_{\perp }=0$) with $k_{z}$ values where the spacing of the intracycle pattern is equal to
(multiple of or one over a multiple of) the intercycle spacing.
The position of the center of the side rings follows a scale law depending on the
Keldysh parameter $\gamma$.

\begin{acknowledgments}
We thank X-M. Tong for sending TDSE data of Fig. 1b. Work supported by CONICET PIP0386, PICT-2016-0296 and PICT-2014-2363 of ANPCyT (Argentina),
and the University of Buenos Aires (UBACyT 20020130100617BA).
\end{acknowledgments}

\bibliography{biblio-diego}

\end{document}